\documentclass[twocolumn, showpacs,preprintnumbers,amsmath,amssymb]{revtex4}
\usepackage{amssymb}
\usepackage{xcolor}
\usepackage{graphicx}% Include figure files
\usepackage{dcolumn}% Align table columns on decimal point
\usepackage{bm}% bold math
\usepackage{multirow}
\usepackage{booktabs}
\usepackage{natbib}
\usepackage{soul}
\usepackage{amssymb}
\usepackage{amsmath}	% Advanced maths commands
\usepackage{color}
\usepackage{cases}

\begin{document}
\preprint{APS/123-QED}

\title{Electron collision studies on the CH$_2^+$ molecular ion}

\author{K. Chakrabarti$^{1}$}\email[]{kkch@eth.net}
\author{J. Zs Mezei$^{2,3}$}\email[]{mezei.zsolt@atomki.hu}
\author{I. F. Schneider$^{3}$}\email[]{ioan.schneider@univ-lehavre.fr}
\author{J. Tennyson$^{3,4}$}\email[]{j.tennyson@ucl.ac.uk}
\affiliation{$^1$Department of Mathematics, Scottish Church College, 700006 Kolkata, India}%
\affiliation{$^{2}$HUN-REN Institute for Nuclear Research (ATOMKI), H-4001 Debrecen, Hungary}%
\affiliation{$^{3}$LOMC-UMR6294, CNRS, Universit\'e Le Havre Normandie, 76600 Le Havre, France}%
\affiliation{$^{4}$Department of Physics and Astronomy, University College London, WC1E 6BT London, UK}%
\date{\today}

\begin{abstract}
Calculations are performed for electron collision with the methylene molecular ion CH$_2^+$ in its bent equilibrium geometry, with the goal to obtain cross sections for electron impact excitation and dissociation. The polyatomic version of the UK molecular R-matrix codes was used to perform an initial configuration-interaction calculation on the doublet and quartet states of the CH$_2^+$ ion. Subsequently, scattering calculations are performed to obtain electron impact electronic excitation and dissociation cross sections and, additionally, the bound states of the CH$_2$ molecule and Feshbach resonances in the $e$-CH$_2^+$ system. 
\end{abstract}

%\keywords{coupled-channel, optical shielding, KCs}%Use showkeys class option if keyword

\maketitle

\section{Introduction}
Many low temperature plasma environments have hydrocarbon molecular ions as 
important constituents. Collision of electrons with molecules and their ions 
in these environments are important processes that play a fundamental role in
initiating chemistry, particle balance and transport. For example,
although the present tendency in the magnetically controlled International Thermonuclear Fusion Reactor type fusion devices is 
to coat the reactor walls with beryllium or tungsten, hydrocarbon ions, in 
particular the methylene ion CH$_2^+$, are found in the edge and divertor plasmas 
of  fusion devices  which operate with graphite as plasma 
facing material \cite{McLean05}. Another important context is that of the dusty plasmas, the CH$_2$ and 
CH$_2^+$ being species involved in the chain of reactions resulting in the growths of the nano and micro-particles. In these situations, the cross sections for different 
electron induced process are necessary to model the plasma flow (see for example \cite{Reiter10}), in particular 
electron impact dissociation and dissociative ionisation: 
\begin{equation}\label{eqn:react1}
e + \mathrm{CH_2^+} \rightarrow \left\{ \begin{array}{ll}
\mathrm{C^+} + \mathrm{H_2}+e\\
\mathrm{CH^+} + \mathrm{H}+e \end{array} \right.
\end{equation}
\begin{equation}\label{eqn:react2}
e + \mathrm{CH_2^+} \rightarrow \left\{ \begin{array}{lll}
e\; +\; \mathrm{CH^+} + \mathrm{H^+} + e\\
e\; +\; \mathrm{C^+} + \mathrm{H^+} + \mathrm{H} +e\\
e\; +\; \mathrm{C^+} + \mathrm{H{_2}^+}+e \\ 
e\; +\; \mathrm{C^+} + \mathrm{H^+} + \mathrm{H^+} + 2e \end{array} \right .
\end{equation}
These processes,  leading to the destruction of the  CH$_2^+$ ions and to the  C 
atom production \cite{Janev02a,Janev02b}, are highly significant for 
understanding the carbon redeposition.

In the interstellar medium (ISM), CH$_2^+$ ions are synthesised in gas phase by
collision of C$^+$ ions with hydrogen \cite{Wak10} and through hydrogen 
abstraction by CH$^+$. On the other hand, CH$_2^+$ ions can be removed by reactions
such as (\ref{eqn:react1}), (\ref{eqn:react2}) and by dissociative recombination
(DR), 
\begin{equation}
e + \mathrm{CH_2^+} \rightarrow \left\{ \begin{array}{lll}
\mathrm{C + H_2}\\
\mathrm{CH + H}\\
\mathrm{C + H + H} \end{array} \right.
\end{equation}
which is known to proceed via Feshbach resonances \cite{Larson98}.

There is a considerable literature  on CH$_y^+$ hydrocarbon ions, and CH$_2^+$ 
ions in particular, in the context of synthesis of hydrocarbons in the ISM 
\cite{Dishoeck96,Dishoeck06,Wak10,Puglisi18,IdB19}. Significant work on  
CH$_y^+$ hydrocarbon ions have also been done relevant to plasmas for fusion 
\cite{Janev02a,Janev02b,Vane07,Lecointre09,Reiter10}. These works mainly focus 
on different electron impact cross sections relevant for plasma modeling.

Molecular structure calculations on CH$_2^+$ have been reported by several authors
\cite{JT83,TP91,KJB94,Li15,Guo18,Ma21}. \cite{TP91} obtained bending potential 
energy curves (PEC) and vertical excitation energies of CH$_2^+$ for its bent $C_{2 
v}$ and its linear $D_{\infty h}$ configurations. Accurate global potential 
energy surfaces (PES) were reported by \cite{Guo18} and \cite{Li15}. Apart from these, 
there are also many spectroscopic studies on the rovibrational states of CH$_2^+$ 
\cite{Ross92,Bunker01,Jensen02,Will03,Wang13}.

In a number of earlier studies 
\cite{Chakrabarti17,Chakrabarti18,Chakrabarti19a,Chakrabarti20} we have
studied electron collision with the CH molecule and its positive ion CH$^+$ in
considerable detail. In these works, we  not only computed cross sections 
for 
different electronic processes, but also identified many new neutral valence
states of CH that are relevant for the DR. The present article aims to continue 
and extend our work to more complex hydrocarbon ions. 

\section{R-matrix Calculations}
\subsection{R-matrix method}
The R-matrix method, described in detail by \cite{Tennyson10} and
\cite{Burke11}, and its implementation in the polyatomic version of the UK 
molecular R-matrix codes (UKRmol) \cite{jt518} is used in the present work. The method employs a 
division of  space into an inner region, a sphere of radius $a$ 
(chosen to be 10~a$_0$ in this work) called the $R$-matrix sphere whose purpose is 
to include within it all short range interactions, and an outer region exterior 
to this sphere which contains all the long range interactions. This division allows the treatment of the short range and, the more complicated, 
long range interactions separately using different techniques \cite{Tennyson10}.

In the inner region, the wave function of the target (here CH$_2^+$) and a 
single continuum electron, together having $N+1$ electrons, is taken as 
\begin{eqnarray}\label{eq:cc}
\Psi_k &=& {\cal A} \sum_{i,j} a_{i,j,k} \Phi_i(1,\ldots,N)
F_{i,j}(N+1) + \nonumber \\
&+&\sum_i b_{i,k} \chi_i(1,\ldots,N+1) \;,
\end{eqnarray}
where $\mathcal{A}$ is an antisymmetrisation operator, $\Phi_i(1,\ldots,N)$ is 
the wave function of the $N$ electron target and $F_{i,j}(N+1)$ are continuum 
orbitals. The functions $\chi_i(1,\ldots,N+1)$ in the last term are square 
integrable functions, called $L^2$ functions, are constructed by allowing the 
projectile electron to enter the target complete active space (CAS) and are 
included to take into account electron correlations and  polarization of the 
target in presence of the projectile electrons. The coefficients 
$a_{i,j,k}$ and $b_{i,k}$ are obtained by diagonalizing the inner region 
Hamiltonian.

The inner region wave function $\Psi_k$ is then used with appropriate boundary 
conditions to obtain scattering information, the details of which are given in 
the following subsections.

\subsection{Target calculations}
We  used the cc-pVTZ Gaussian basis sets \cite{BPP19} centered on the C and 
H atoms to represent the target orbitals. These not only gave reasonably good 
target vertical excitation energies but also allows  the inner region 
calculation to remain manageable with respect to computational resources. 

The X~$^2A_1$ ground state of CH$_2^+$ is known to be bent in C$_{\rm 2v}$ symmetry 
and has the electronic 
configuration $(1a_1)^2 (2a_1)^2 (1b_2)^2 (3a_1)^1$ \cite{Pople87,Graber93}. In 
this work all calculations are reported at the equilibrium C-H bond length 
2.066~a$_0$ and the H-C-H bond angle 140.1$^\circ$ taken from \cite{TP91}, 
which are very close to those obtained by more sophisticated coupled cluster 
calculations using large basis sets \cite{NRB02}. An initial Hartree-Fock (HF) 
calculation was first performed on the X~$^2A_1$ ground state of CH$_2^+$. The HF 
orbitals were then used in a configuration interaction (CI)  calculation.

We considered two target models. In both models we kept two electrons frozen in
the $1a_1$ orbitals while the remaining five electrons were distributed in the 
CAS. In the first model (M1), the complete active space was defined by 
$(1a_1-6a_1, 1b_1-4b_1,1b_2-4b_2)$, while in the second (M2) the CAS was chosen 
to be bigger with an additional 1$a_2$ orbital, namely $(1a_1-8a_1, 1b_1-5b_1, 
1b_2-5b_2, 1a_2)$. Table \ref{tab:Target} shows the comparison of the vertical 
excitation energies (VEE) form the X~$^2A_1$ ground state of CH$_2^+$ to the first 
$12$ low lying excited states. Although the second model M2 appears to produce 
VEEs slightly in better agreement with the theoretical results of 
\cite{TP91} and \cite{Osmann99}, calculations with this model 
required much longer time compared to the model M1. We therefore chose the model 
M1 for subsequent calculations as it was computationally more efficient.

From Table \ref{tab:Target}, we see that apart from the VEE of the 4~$^2A_1$ 
state, which appears too high compared to \cite{TP91}, all other vertical 
excitation energies obtained by the target model M1 are in reasonably good 
agreement  with the results of \cite{TP91} and \cite{Osmann99}. 
Moreover, our dipole moment for the CH$_2^+$ ground state with model M1 is 0.701 D 
which compares perfectly with the value 0.701 D obtained by \cite{NRB02} 
using coupled cluster calculation. The model M1 therefore provides a good 
description of the target for subsequent scattering calculations.

\begin{table}[t]
\caption{\label{tab:Target} 
Comparison of the vertical excitation energies (in eV) from the X~$^2A_1$ ground 
state to 12 low lying excited states of CH$_2^+$ for different target models. The 
target models used are the following:\\
M1: $(1a_1)^2 (1a_1-6a_1, 1b_1-4b_1,1b_2-4b_2)^5$\\
M2: $(1a_1)^2 (1a_1-8a_1, 1b_1-5b_1, 1b_2-5b_2, 1a_2)^5$}
\begin{tabular} {lrrrr} 
\hline
Target state & M1 & M2 & Theory$^a$ & Theory$^b$\\
\hline
X~$^2A_1$& 0$^c$ & 0$^d$ & 0$^e$ & 0   \\
1~$^2B_1$ & 0.92 & 0.81 & 0.84 & 0.83 \\
1~$^4A_2$ & 5.24 & 5.07 & & \\
1~$^2A_2$ & 6.93 & 6.81 & 6.81 & 6.92\\
1~$^2B_2$ & 7.46 & 7.36 & 7.60 & 7.8\\ 
2~$^2A_2$ & 7.88 & 7.60 & 7.25 & 7.26\\ 
1~$^4B_1$ & 9.73 & 9.58 & & \\
2~$^2B_2$ & 9.48 & 9.17 & 9.25 & 9.6 \\
2~$^2A_1$ & 11.17 & 10.50 & 10.44 & 11.1\\
3~$^2A_1$ & 13.52 & 13.04 & 12.81 & \\ 
2~$^2B_1$ & 13.47 & 13.21 & 13.14 & 13.9\\
4~$^2A_1$ & 14.89 & 14.76 & 13.40 & \\
3~$^2B_2$ & 13.22 & 13.23 & 13.53 & \\
\hline
\end{tabular}\\
$^a$\cite{TP91}\\
$^b$\cite{Osmann99}\\
$^c$Absolute energy -38.61306904 Hartree\\
$^d$Absolute energy -38.63754516 Hartree\\
$^e$Absolute energy -38.705744  Hartree
\end{table}

\subsection{Scattering calculations}
For the scattering calculations, we used $8~a_1, 6~b_1, 6~b_2 \textrm{ and } 2~a_2$ target
orbitals allowing $2$ virtual orbitals for each symmetry. Since the target 
CH$_2^+$ is a positive ion, the continuum functions were represented by Coulomb 
functions which were obtained as a solution of the radial Coulomb  equation for 
an isotropic Coulomb potential, and the solutions with $l\le4$, and energy 
eigenvalue $\le 5$ Ryd were retained in the calculation. The Coulomb functions 
were then fitted to GTOs using the procedure outlined in \cite{Faure02}.

The target and continuum orbitals must all be orthogonal to each other. To 
ensure the orthogonality, the target and continuum orbitals were first 
individually Schmidt orthogonalized and finally the full set of target and 
continuum orbitals were symmetric orthogonalized, retaining only those orbitals 
for which the eigenvalue of their overlap matrix was less than a deletion 
threshold, chosen here to be $5\times 10^{-5}$. The deletion threshold depends 
on the R-matrix radius and was adjusted to ensure that there was no linear dependence.

An R-matrix was built at the boundary of the inner 
and outer region from the inner region solutions (Eq.(\ref{eq:cc})). The $R$-matrix were then propagated to an asymptotic distance $R_{asy} =70\,$a$_0$ in a potential which included the Coulomb 
potential and the dipole and quadrupole potentials of the target, where they 
were then matched to asymptotic functions obtained from a Gailitis expansion 
\cite{Noble84}. This matching procedure yields the $K$-matrix from which all 
scattering observables can be obtained. A different route, however, is 
followed for obtaining bound states, as is outlined below.

For bound states, the R-matrix and wave functions were propagated using Runge-
Kutta-Nystrom method \cite{Zhang11} to an asymptotic distance $R_{asy}=50\,$a$_0$
in a potential which included the Coulomb potential and the dipole and quadrupole potentials of target, and were matched to exponentially decreasing asymptotic functions \cite{Noble84}. A searching algorithm \cite{Sarpal91} over a non-linear quantum defect based grid 
\cite{Rabadan96} was then used to find the bound states as roots of a 
determinant $\mathcal{B}(E)$ dependent on the energy. The details of the 
method are omitted and can be found in \cite{Sarpal91}.

\begin{table}[t]
\caption{ Comparison of the vertical excitation energies (in 
eV) from the ground X~$^3B_1$ state of CH$_2$ to some of its low lying excited 
states.}
\label{tab:ch2ve}
\begin{tabular} {lcccc} 
\hline
CH$_2$ state & This work & CI$^a$ & Romelt$^b$ & Yamaguchi$^c$\\
\hline
X~$^3B_1$ & 0.0$^d$  & 0.0$^e$ & 0.0$^f$ & 0.0\\ 
1~$^1A_1$ & 1.16 & 0.995 & 1.14 & \\
$^1B_1$   & 1.74 & 1.86 & 1.63  & \\
2~$^1A_1$ & 3.24 & 3.31 & 3.38  & \\
$^3A_1$   & 6.21 & 7.67 & 6.37& \\
$^3B_2$   & 7.35 & 8.05 & 7.59  & 7.86\\
$^1B_2$   & 7.25 & 9.25 & 7.63  & 7.75\\
$^3A_2$   & 7.63 & 7.47 & 7.57  & 7.23\\
$^1A_2$   & 8.33 & 8.39 & 8.46 & \\
\hline
\end{tabular}\\
$^a$CI calculation done at CH$_2$ equilibrium (C-H bond length 2.0314~a$_0$ and 
H-C-H bond angle 133.8$^\circ$) using $(1a_1)^2~(2a_1-6a_1, 
1b_1-4b_1,1b_2-4b_2)^6$ CAS-CI model.\\
$^b$\cite{Romelt81}\\
$^c$\cite{Yama97}\\
$^{d,e,f}$ Absolute energies of the ground states are respectively      
$-38.985233^d$ Hartree, $-38.966950^e$ Hartree and $-39.06034^f$ 
Hartree.\\
\end{table}
\section{Results}
All scattering calculations in this work were performed at a single geometry, 
namely the equilibrium geometry of the CH$_2^+$ target ion, for bound states of 
the CH$_2$ molecule, Feshbach resonances in the $e+$CH$_2^+$ system and cross 
sections for elastic scattering and electronic excitations. An 11 state 
scattering model including three $^2A_1, \textrm{ three }^2B_1, 
\textrm{ three } ^2B_1$ and two $^2A_2$ target states in the close coupling 
expansion Eq. (\ref{eq:cc}) were used for the singlet $e+ $CH$_2^+$ scattering 
close-coupling expansion, while a 15 state model which included the same 11 
doublet states as in the singlet state model, together with the lowest each 
of the $^4A_1, ^4B_1, ^4B_2$, and $^4A_2$ states was used in Eq. (\ref{eq:cc}) 
for triplet symmetry close-coupling expansion. As will be shown below, this 
procedure provides a  reliable scattering model which was tested by calculating 
the bound states of CH$_2$.

\begin{figure}[t]
\begin{center}
\includegraphics[width=0.9\columnwidth]{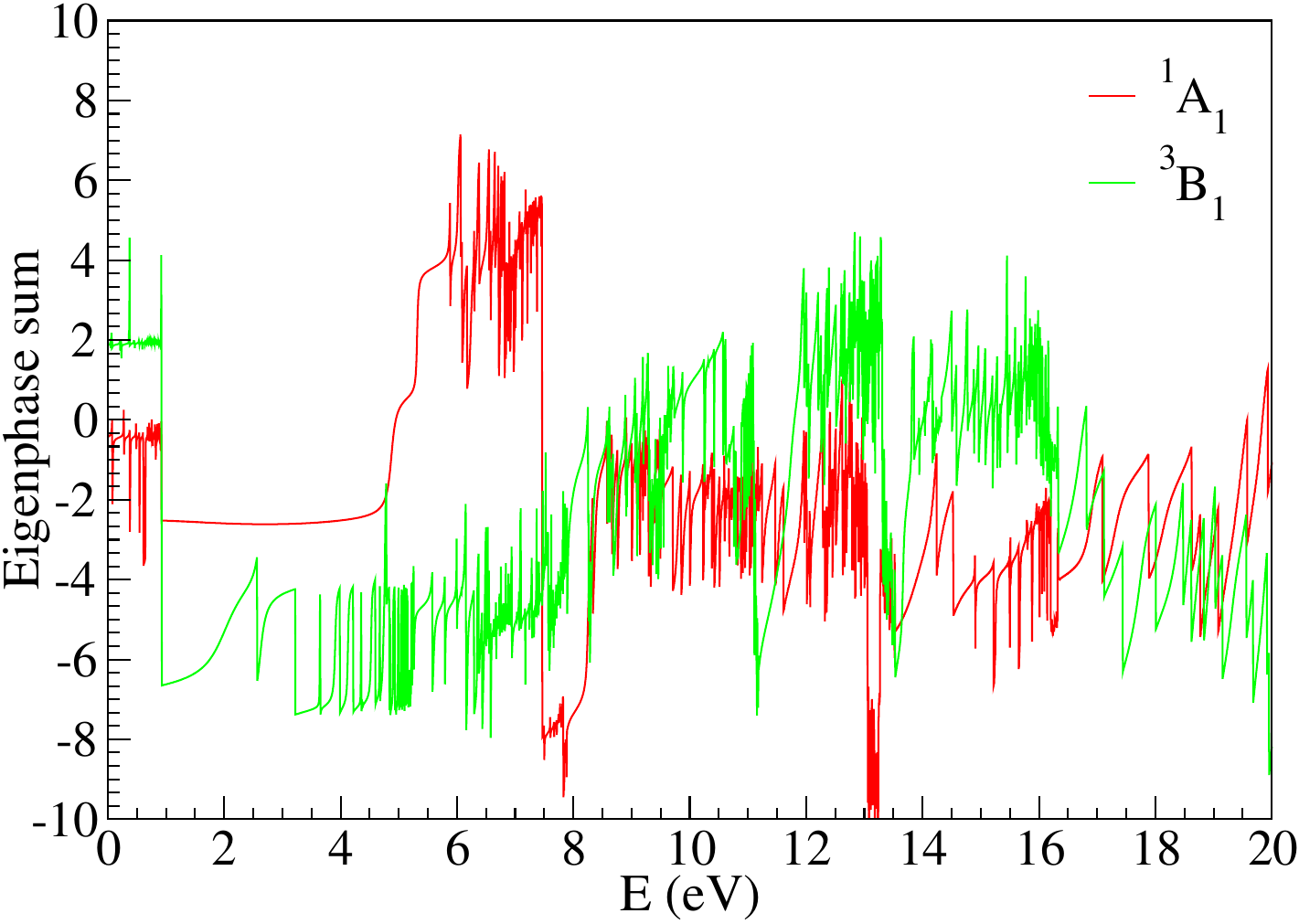}
\end{center}
\caption{Eigenphase sums for the overall $e +$ CH${_2}^+$ symmetries
$^1A_1$ and $^3B_1$.
\label{fig:eigenp}
}
\end{figure}

\subsection{Bound states}
Table \ref{tab:ch2ve} shows the vertical excitation energies for the bound 
states of CH$_2$ form its X~$^3B_1$ ground state to few of its low lying excited 
states. For a comparison, we also did a quantum chemistry-style CI calculation 
on CH$_2$ at its equilibrium geometry (C-H bond length 2.0314~a$_0$ and H-C-H bond 
angle 133.8$^\circ$) using a $(1a_1)^2~(2-6a_1, 1-4b_1, 1-4b_2)^6$ CAS-CI model. 
The vertical excitation energies are then compared with the multi reference 
double excitation (MRD-CI) calculation of \cite{Romelt81} and the coupled 
cluster results of \cite{Yama97}. Form Table \ref{tab:ch2ve}, it is clear 
that the vertical excitation energies obtained by the R-matrix method are in very 
good agreement with all others. The fact that the calculated bound state 
energies are consistent and accurate enough indicates that our scattering model 
is reasonably good.

\begin{figure*}[t]
\begin{center}
\includegraphics[width=0.42\textwidth]{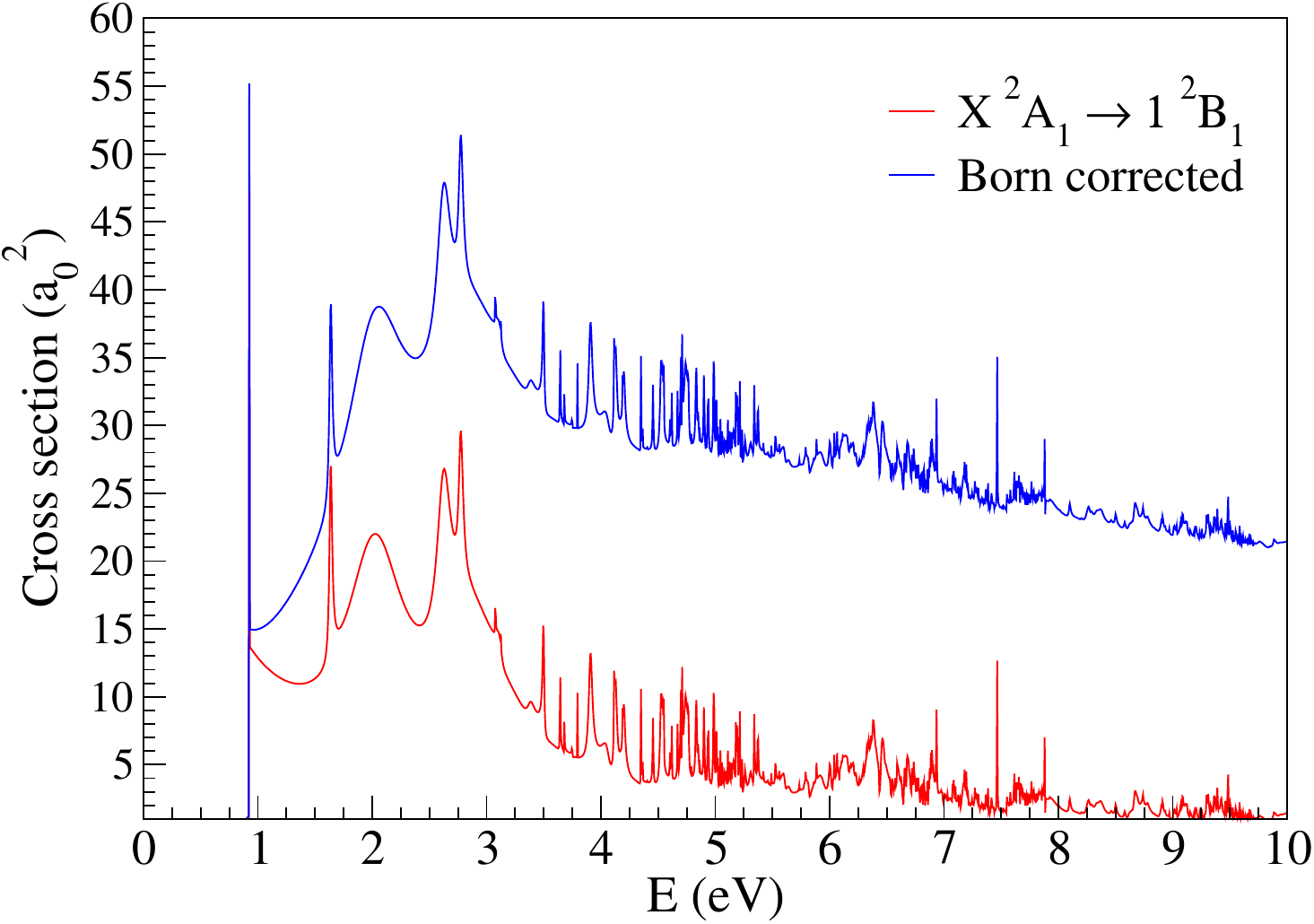}
\includegraphics[width=0.42\textwidth]{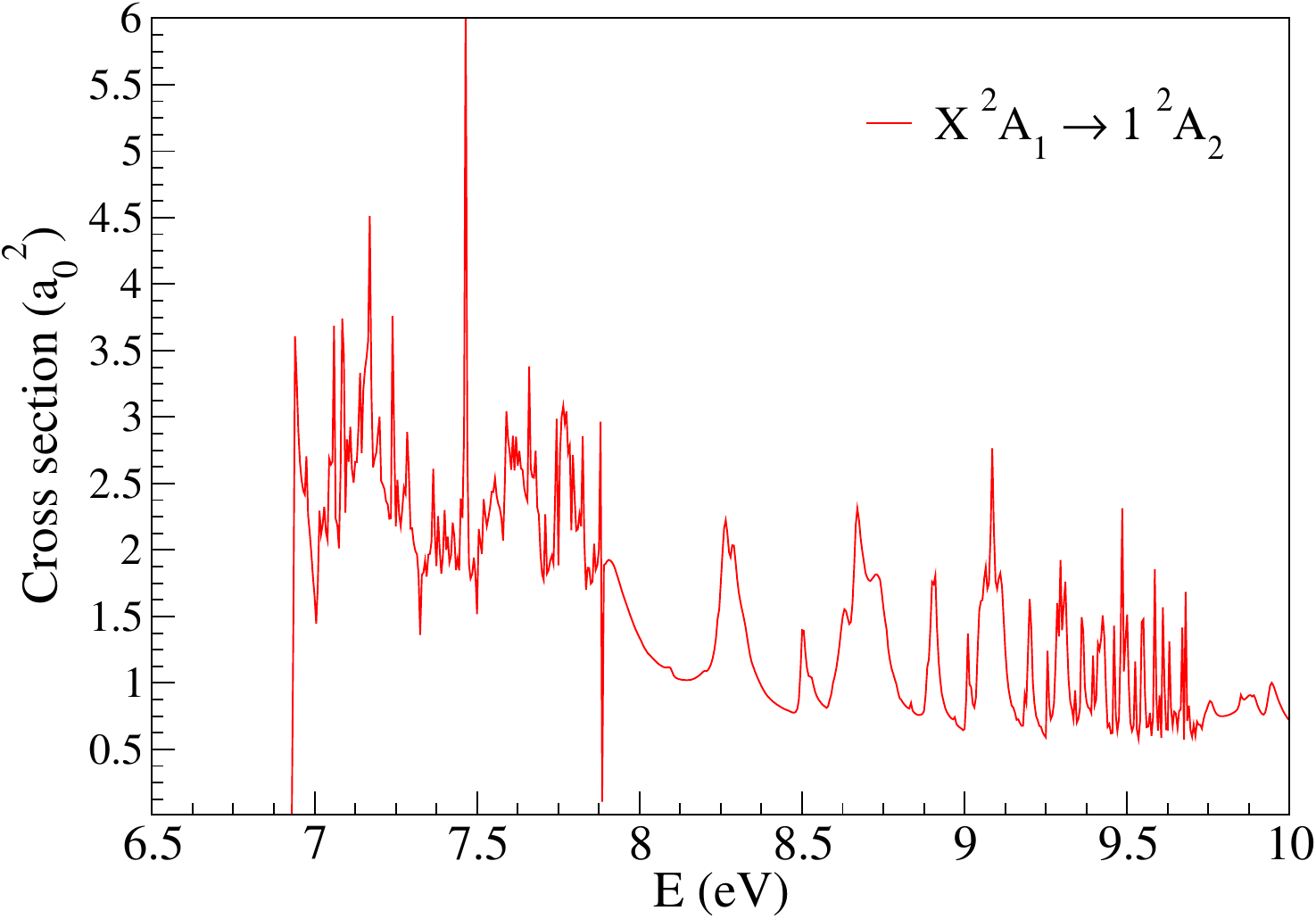}
\includegraphics[width=0.42\textwidth]{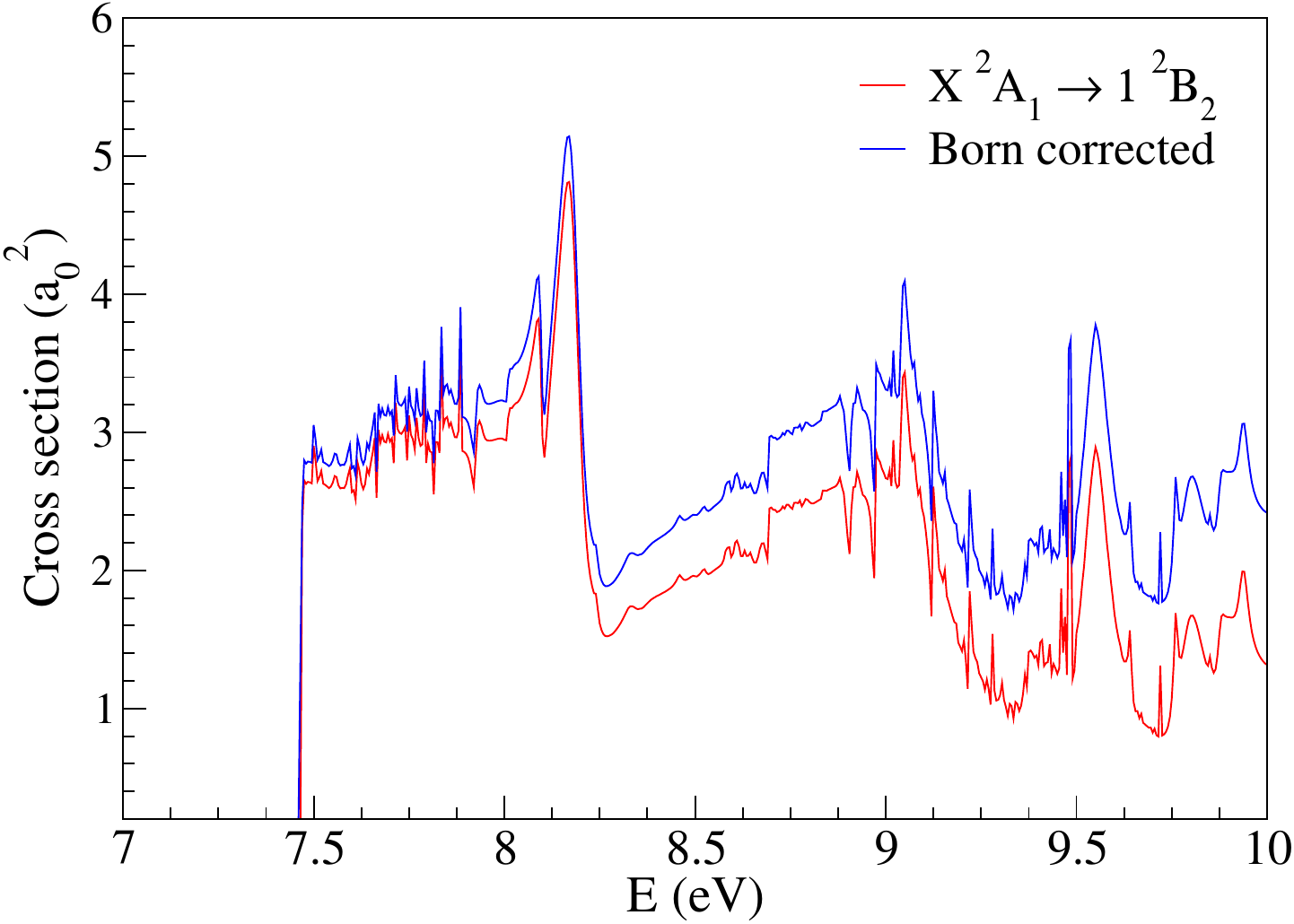}
\includegraphics[width=0.42\textwidth]{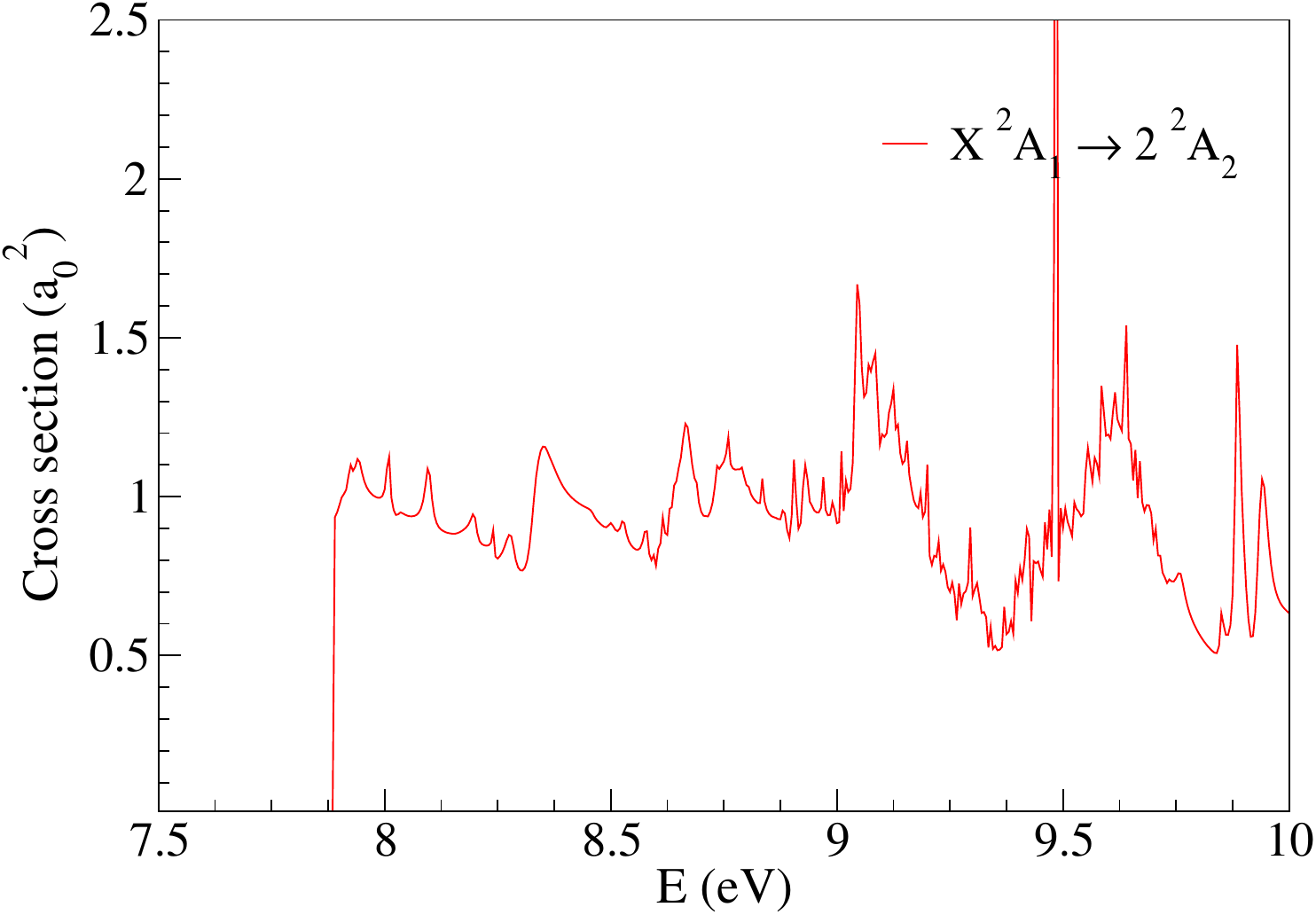}
\end{center}
\caption{Cross sections for electronic excitation form the $X~^2A_1$ ground 
state to the first four excited states of doublet symmetry as given in Table 
\ref{tab:Target} and indicated in each panel.
\label{fig:excitation}
}
\end{figure*}

\subsection{Resonances at equilibrium}
For resonance calculation, the R-matrix was propagated to 70~a$_0$. Resonances were 
detected by the characteristic change in sign of the second derivative of the 
eigenphase sum $\delta(E)$ given by
\begin{equation}
\delta(E)=\sum_i \arctan(K_{ii}),
\end{equation}
where $K_{ii}$ are the diagonal elements of the $K$ matrix. They were then 
fitted to a Breit-Wigner profile \cite{Tennyson84} with an energy grid 0.005~eV 
to obtain the resonance energy $E$ and width $\Gamma$.

\begin{table}[t]
\caption{Resonance positions  and widths (in Ryd) and effective
quantum numbers at the CH$_2^+$ equilibrium for states of $^3B_1$ and $^1A_1$ 
symmetry of the $e$-CH$_2^+$ system below the first two CH$_2^+$ excited states. 
Numbers within brackets indicate power of 10.}
\label{tab:reson}
\begin{tabular}{cccccc}
\hline 
Position & Width & $\nu$ & Position & Width & $\nu$\\
\hline
$^3B_1$ symmetry\\
\multicolumn{3}{c}{Below $1~^2B_1$ state}&
\multicolumn{3}{c}{Below $~^4A_2$ state}\\
0.3558(-02)& 0.2400(-04)& 3.9466 & 0.1925& 0.1075(-01) & 2.2792\\
0.4870(-02)& 0.5833(-04)& 3.9875 & 0.2678& 0.3334(-03) & 2.9201\\
0.1567(-01)& 0.4765(-04)& 4.3815 & 0.2872& 0.2789(-02) & 3.1980\\
0.1718(-01)& 0.6172(-04)& 4.4465 & 0.3034& 0.1910(-03) & 3.5000\\
0.2685(-01)& 0.1503(-04)& 4.9440 & 0.3197& 0.2426(-03) & 3.9143\\
\hline 
$^1A_1$ symmetry\\
\multicolumn{3}{c}{Below $1~^2B_1$ state}&
\multicolumn{3}{c}{Below $1~^2A_2$ state}\\
0.2034(-01)& 0.2198(-03) & 4.5925 & 0.3582 & 0.9116(-02) &2.5716\\
0.3582(-01)& 0.1225(-03) & 5.5953 & 0.3907 & 0.2355(-02) &2.9015\\
0.4478(-01)& 0.7510(-04) & 6.5969 & 0.4323 & 0.1280(-02) &3.6009\\
0.5043(-01)& 0.4929(-04) & 7.5979 & 0.4424 & 0.2339(-02) &3.8615\\
0.5423(-01)& 0.3407(-04) & 8.5985 & 0.4492 & 0.3074(-02) &4.0762\\
\hline
\end{tabular}
\end{table}

Figure \ref{fig:eigenp} shows the plot of two typical eigenphase sums for 
$^1A_1$ and $^3B_1$ symmetries. Characteristic of electron collision with ions, 
the figure shows numerous resonances some of which are tabulated in Table 
\ref{tab:reson} according to their parent state. As seen from the table, many of 
the resonances appear to be in Rydberg series which can be identified by their 
effective quantum numbers. The relatively small gap between the ground
and first excited electronic state means that the resonances start with
relatively high effective quantum numbers, $\nu \approx 4$, and 
then closely space.

A full set of fits to the resonances covering all overall scattering
symmetries is given in the supplementary material.

\subsection{Electron impact excitation}
Our calculated cross sections for electron impact excitation of CH$_2^+$ from the 
ground state to four of its lowest doublet states are shown in Figure 
\ref{fig:excitation}. As in the elastic cross section, the excitation cross 
sections show highly resonant behaviour due to temporary captures into resonant 
states. Particularly, the X~$^2A_1 \rightarrow~1~^2B_1$ excitation cross section 
shows evidence of a large resonance near threshold. Referring to Figure 
\ref{fig:eigenp}, this is likely due to the large $^3B_1$ resonance near 1~eV as 
seen from the plot of the $^3B_1$ eigenphase sum. Generally, the excitation 
cross sections decrease with increasing incident energy and the cross sections 
from the higher lying excited states are much smaller.

\begin{figure}[t]
\begin{center}
\includegraphics[width=0.9\columnwidth]{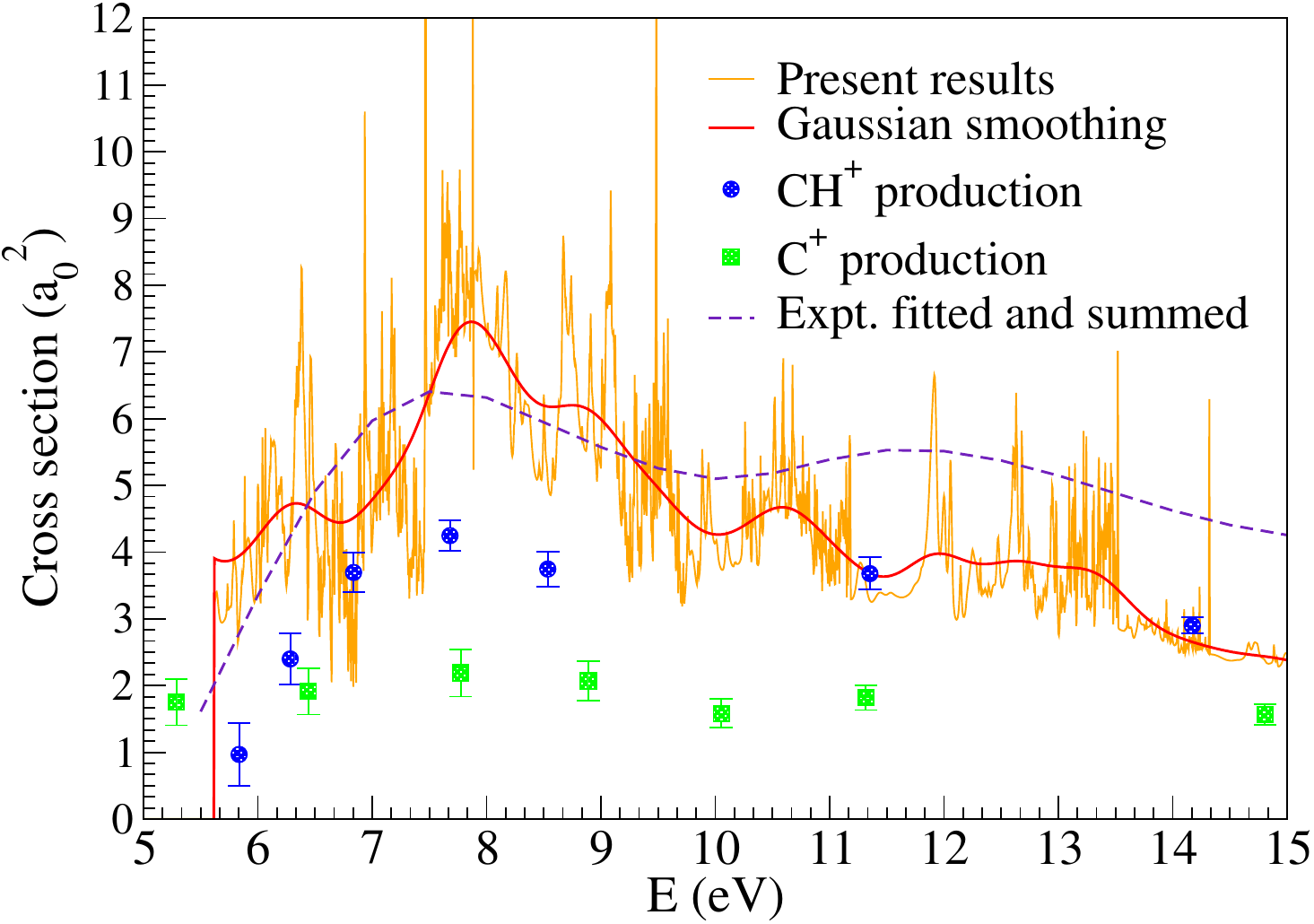}
\end{center}
\caption{Cross section for the electron impact dissociation of the CH$_2^+$ ion.
Thin curve: present R-matrix results (see text). Thick (red) curve: present result after smoothing with a Gaussian function. Green circles
with error bars: experimental results for C$^+$ ion production 
\cite{Vane07}. Blue circles with error bars: experimental results 
for CH$^+$ ion production \cite{Vane07}. Dashed curve: 
experimental results for C$^+$ and CH$^+$ ion production summed after suitable 
fitting (see text).
\label{fig:diss}
}
\end{figure}

\subsection{Electron impact dissociation}
Experimental cross sections for electron impact dissociation were obtained 
by \cite{Vane07} for the production of CH$^+$ and C$^+$ 
fragments. Although model calculations exist for the electron impact 
dissociation of CH$_2^+$ ions (see for example \cite{Reiter10}), to the best 
of our knowledge, the 2007 experiments have never been described by any \textit{ab 
initio} calculation.

The dissociation of CH$_2^+$ into the lowest dissociation channels, namely 
$e +$ CH$^+$ + H and $e +$ C$^+$ + H$_2$, proceed via direct dissociative 
excitation and have thresholds 6.08 eV and 5.62 eV respectively \cite{Vane07}. 
In deriving the dissociation cross section, we assumed that electronic 
excitations to all states above the respective dissociation thresholds lead to 
dissociation. In our calculation, we have included the states 1~$^2B_1$,  
1~$^2B_2$, 1~$^2A_2$, and the 2~$^2A_2$ excited states all of which lie close 
to one another at the CH$_2^+$ equilibrium.  Moreover, except the 1~$^2B_1$ 
state,  which has 
both valence and Rydberg character, all the other 
three are of valence character at equilibrium (see for example \cite{TP91}) and 
hence are likely to dissociate to the e+C$^+$+H$_2$ (5.62 eV) or the  
e+CH$^+$+H (6.08 eV) dissociation limits, which are the most relevant in the
energy range considered. Since the dissociation channels cannot 
be separated in our calculations, our cross section in Figure \ref{fig:diss} 
represents a sum over these channels. For better comparison, we have also shown 
our cross sections after smoothing by using a Gaussian function. The 
experimental data from \cite{Vane07} are available for each of the 
dissociation channels mentioned above. The energy grid of the experimental cross sections for these two dissociation channels, however, are neither same nor 
uniform. Therefore, to sum the experimental data, we first spline interpolated 
each set over the same and uniform energy grid. This allowed us to sum the cross section for the experimental data. In Figure \ref{fig:diss}, this sum is shown 
as the dashed line and it agree fairly well with our computed cross section. In particular, the peak near 8 eV agrees quite well with the spline interpolated 
summed experimental curve. However, we note that since the experimental data 
is non uniformly spaced, the interpolated curve may not often reflect the actual trend. For example, we suspect the agreement of the interpolated curve with our calculation would have been much better between 11 eV - 15 eV had there been 
more experimental points available in this region. Similarly, below 6 eV 
the interpolated experimental curve, appears to diverge for our Gaussian fitted
curve. This is a result of the fitting procedure, the interpolated experimental 
curve actually follows the trend of our raw cross section curve.

\section{Conclusion}
As mentioned in the introduction, CH$_2^+$ is a very important constituent in 
low temperature plasma environments. However, despite its importance, electron 
collisions studies on CH$_2^+$ have been rare. In fact, we could not find any {\it ab initio} 
calculation for the cross sections included in the present work. The only cross 
section result available, seems to be the total ionisation cross section of 
CH$_2^+$ calculated within the Binary-Encounter-Bethe (BEB) model by 
\cite{Irikura02}.

In this work, we have presented a  reliable set of cross 
sections for the electronic excitation, and electron impact 
dissociation of the CH$_2^+$ ion. In fact, none of these cross sections have ever been 
reported before by any {\it ab initio} calculation. Additionally, we have also calculated and have 
given the position and width for some of the Feshbach resonances in the e-CH$_2^+$
system. These Feshbach resonances, as is well known, are the routes to 
dissociative recombination of the CH$_2^+$ ion. However, its treatment requires a 
more comprehensive calculation of the resonance energies, widths and the 
potential energy surfaces of the CH$_2^+$ states. This is the subject of an ongoing project.
%%%%%%%%%%%%%%%%%%%%%%%%%%%%%%%%%%%%%%%%%%%%%%%%%%
\section*{Acknowledgements}
The authors acknowledge support from F\'ed\'eration de Recherche Fusion par Confinement Mag\'etique (CNRS, CEA and Eurofusion), La R\'egion Normandie, FEDER and LabEx EMC3 via the projects PTOLEMEE, Bioengine, the Institute for  Energy, Propulsion and Environment (FR-IEPE),and ERASMUS-plus conventions between Université Le Havre Normandie and University College London. We are indebted to Agence Nationale de la Recherche (ANR) via the project MONA, Centre National de la Recherche Scientifique via the Programme National ’Physique et Chimie du Milieu Interstellaire’ (PCMI). JZsM thanks the financial support of the National Research, Development and Innovation Fund of Hungary (NKFIH), under the K18 and FK19 funding schemes with project no. K128621 and FK132989.
JZsM and IFS are grateful for the support of the NKFIH--2019-2.1.11-TÉT-2020-00100 and Campus France-Programme Hubert Curien--BALATON--46909PM projects.
%\onecolumn
\section*{Data availability}
Upon a reasonable request, the data supporting this article will be provided by the corresponding author.

\end{document}